\documentclass[12pt]{article}
\usepackage{geometry}                
\usepackage{graphicx}
\usepackage{amssymb}
\usepackage{amsmath}
\usepackage{epstopdf}
\def\ba{\begin{array}}
\def\ea{\end{array}}

\def\dalemb#1#2{{\vbox{\hrule height .#2pt
        \hbox{\vrule width.#2pt height#1pt \kern#1pt
                \vrule width.#2pt}
        \hrule height.#2pt}}}

\def\Im{{{\frak{Im}}}}
\def\Re{{{\frak{Re}}}}

\def\Op{{\mathcal{O}}}

 \newcommand{\be}{\begin{equation}}
\newcommand{\ee}{\end{equation}}
 \newcommand{\bal}{\begin{align}}
  \newcommand{\eal}{\end{align}}
 \newcommand{\ben}{\begin{equation*}}
\newcommand{\een}{\end{equation*}}
\newcommand{\bea}{\begin{eqnarray}}
\newcommand{\eea}{\end{eqnarray}}
\newcommand{\bean}{\begin{eqnarray*}}
\newcommand{\eean}{\end{eqnarray*}}
\newcommand{\bes}{\begin{subequations}}
\newcommand{\ees}{\end{subequations}}

\begin{document}

\begin{titlepage}
\bigskip
\rightline{}

\bigskip\bigskip\bigskip\bigskip
\centerline {\Large \bf {Holographic Superconductors }}
\bigskip
\centerline{\Large \bf { with Various Condensates}}
\bigskip\bigskip
\bigskip\bigskip

\centerline{\large  Gary T. Horowitz and Matthew M. Roberts}
\bigskip\bigskip
\centerline{\em Department of Physics, UCSB, Santa Barbara, CA 93106}
\centerline{\em  gary@physics.ucsb.edu, matt@physics.ucsb.edu}
\bigskip\bigskip
\begin{abstract}
We extend earlier treatments of holographic superconductors by studying cases where operators of different dimension condense in both $2+1$ and $3+1$  superconductors. We also compute a correlation length. We find surprising regularities in  quantities such as $\omega_g/T_c$ where $\omega_g$ is the gap in the frequency dependent conductivity. In special cases,   new bound states arise corresponding to vector normal modes of the dual near-extremal black holes.
\end{abstract}
\end{titlepage}

\section{Introduction}

It has recently been shown that a simple Einstein-Maxwell theory coupled to a charged scalar provides a holographically dual description of a superconductor  \cite{Gubser:2008px,Hartnoll:2008vx}. The gravity theory depends on two parameters, the mass $m$ and charge $q$  of the complex scalar field. If $q$ is large, the backreaction of the matter fields on the metric is small and can be neglected. This probe approximation was used in \cite{Hartnoll:2008vx} and will be adopted here as well. (For a discussion of the dependence on $q$, see \cite{us}.) Our main goal is to explore how the properties of the superconductor depend on the scalar mass $m$. This determines the dimension $\lambda$ of the operator that condenses. In addition, the previous studies have focussed on a four dimensional gravity theory which is dual to a $2+1$ superconductor. This is appropriate for many high $T_c$ materials in which the superconductivity is associated with two dimensional planes. Here we also study the five dimensional gravity theory dual to a $3+1$ superconductor. (For other work on holographic superconductors, see \cite{Nakano:2008xc}-\cite{Herzog:2008he}.)

We consider a variety of masses between $m^2=0$ and the Breitenlohner-Freedman bound $m^2_{BF}<0$ marking the boundary of stability for a scalar field in $AdS$.\footnote{We expect theories with $m^2 > 0$ to also be dual to superconductors, but in this case  there is a numerical instability at large radius making it difficult to find solutions with the correct asymptotic  behavior.} We find qualitatively the same behavior for all masses. Below a critical temperature, a charged black hole becomes unstable to developing scalar hair. In the dual theory, this corresponds to the formation of a charged condensate. By perturbing the black hole, one finds that the conductivity $\sigma(\omega)$ in the dual theory has a delta function at $\omega =0$, and a gap $\omega_g$. These are standard properties of s-wave superconductors. There are
special features which arise when $m^2$ saturates the BF bound.  We will see that at low temperature and general $m^2$, $\sigma(\omega)$ has  poles in the complex frequency plane whose locations depend on $m^2$. When $m^2 = m^2_{BF}$ these poles move onto the real axis inside the gap. This produces delta function contributions to  $\Re[\sigma]$ at certain frequencies.  This suggests that the charged quasiparticles are interacting with a strength that depends on $m^2$. When $m^2 = m^2_{BF}$, they form bound states.\footnote{We thank M. Fisher for this interpretation.} From the bulk standpoint, this corresponds to a (vector) quasinormal mode which becomes an actual normal mode at low temperature and $m^2 = m^2_{BF}$.

In the cases we consider, the dimension $\lambda$ of the operator condensing changes by a factor of three for 2+1 superconductors and a factor of two for 3+1 superconductors.  Some quantities change significantly such as the minimum energy $\Delta$ for charged excitations. In particular, the BCS relation $\omega_g = 2\Delta$ which held in \cite{Hartnoll:2008vx} is not true in general. However we have noticed some striking regularities. In particular, in all cases where $\lambda > \lambda_{BF}$, $\omega_g/T_c \approx 8$. This holds to better than 10\% for both the three and four dimensional superconductors, as well as various scalar masses. Since the corresponding BCS value  is $3.5$, this shows that the energy required to break apart the condensate is more than twice the weakly coupled value. The holographic superconductors are indeed strongly coupled. We also find regularities involving the superfluid density $n_s$ and a correlation length $\xi_k$.

In the next section, we describe the gravitational theories we will study and show that  scalar hair forms at low temperatures in all cases. In section 3, we compute the conductivity in the dual theory by considering electromagnetic perturbations of the black holes. Section 4 contains a discussion of other physical quantities, such as the energy gap $\Delta$, the superfluid density $n_s$, and a correlation length $\xi_k$. The final section contains a short summary. In a brief appendix we calculate the conductivity in the normal phase analytically, finding agreement with the numerical results.

\section{Abelian Higgs Model}

We study an Abelian Higgs model in $AdS_{d+1}$:
\be\label{action}
 S=\int d^{d+1}x \sqrt{-g}\left(-\frac{1}{4}F_{\mu \nu}F^{\mu \nu} - |\nabla\Psi-i A\Psi|^2 - m^2|\Psi^2|\right). 
\ee
(Note that we do not include a $\Psi^4$ term.)
If one multiplies this action by $1/q^2$ and sets $\Psi = q \tilde \Psi$, one recovers the standard action for a scalar $\tilde \Psi$ with charge $q$.  In the limit where $q$ is large and $\Psi $ is held fixed, the backreaction on the metric becomes negligible. Following \cite{Hartnoll:2008vx}, we will work in this probe limit.

Our background metric will be the standard  AdS-Schwarzschild black hole
\be
ds^2=-f dt^2+f^{-1}dr^2+r^2dx_i dx^i,\qquad f(r) =r^2(1-r_0^d/r^d),\ee
where we have chosen units such that the AdS radius is unity.
To see the formation of scalar hair, we consider static, translationally invariant solutions to (\ref{action}). Consider the ansatz $\Psi=|\Psi|=\psi$, $A=\phi~dt$ where $\psi,~\phi$ are all functions of $r$ only. This leads to the equations of motion
\bes\label{eom}
\bal
\psi''+\left(\frac{f'}{f}+\frac{d-1}{r}\right)\psi'+\frac{\phi^2}{f^2}\psi-\frac{m^2}{f}\psi=0,\\
\phi''+\frac{d-1}{r}\phi'-\frac{2\psi^2}{f}\phi=0.\label{phieom}\end{align}
\ees
As pointed out in \cite{Gubser:2008px}, the coupling of the scalar to the Maxwell field produces a negative effective mass for $\psi$. Since this term becomes more important at low temperature, we expect an instability toward forming scalar hair. This was indeed seen for $m^2 = -2$ and $d=3$ in \cite{Hartnoll:2008vx}. We now consider $m^2 = 0$ and $m^2 = m^2_{BF} = -9/4$ in $d=3$ and $m^2 = 0,-3,-4$ in $d=4$. Note that the last value of $m^2$ is at the BF bound.
Since (\ref{eom}) are coupled nonlinear equations,  we do not expect to solve them analytically. However, it is straightforward to solve them numerically.

Recall that the asymptotic behavior of a  scalar field in $AdS_{d+1}$ is
\be
\psi = {\psi_-\over r^{\lambda_-}} + {\psi_+\over r^{\lambda_+}}+\cdots\label{asymptscalar}
\ee
 where $\lambda_\pm=\frac{1}{2}(d\pm\sqrt{d^2+4m^2})$. For $m^2\ge-d^2/4+1$, only the $\lambda_+$ mode is normalizable, and so we interperet $\psi_+=\langle\Op\rangle$, where $\Op$ is the operator dual to the scalar field, and $\lambda$ is the dimension of the operator.\footnote{This normalization of $\Op$ differs from that of \cite{Hartnoll:2008vx} by a factor of $\sqrt{2}$.} $\psi_-$ is dual to a source for $\Op$. To have spontaneous symmetry breaking, we wish for the operator to condense without being sourced.  For $-d^2/4\le m^2 < -d^2/4+1$ both modes are normalizable and we have  a choice of quantization. However, stability of $AdS$ dictates that the only stable quantizations which respect $AdS$ symmetries at large radius are ones where $\psi_+$ or $\psi_-$ vanish \cite{Hertog:2004rz}.  When $m^2=-d^2/4=m^2_{BF}$, $\lambda_+=\lambda_-=\lambda_{BF} \equiv d/2$ and there is a logarithmic branch. Such a branch will necessarily introduce an instability unless it is treated as the source \cite{Hubeny:2004cn}.  We have not obtained the Abelian Higgs model from a string theory truncation,  so the dual operator $\Op$ is not known beyond the fact that it is charged under a global $U(1)$ symmetry and has conformal dimension $\lambda$ at strong coupling.
 
 The asymptotic behavior of $\phi$ is
 \be
 \phi = \mu - {\rho\over r^{d-2}}+\cdots\label{asymptphi}
 \ee
 In terms of the dual theory, $\mu$ is the chemical potential and $\rho$ is the charge density.

\begin{figure}\begin{center}
\includegraphics[width=.45\textwidth]{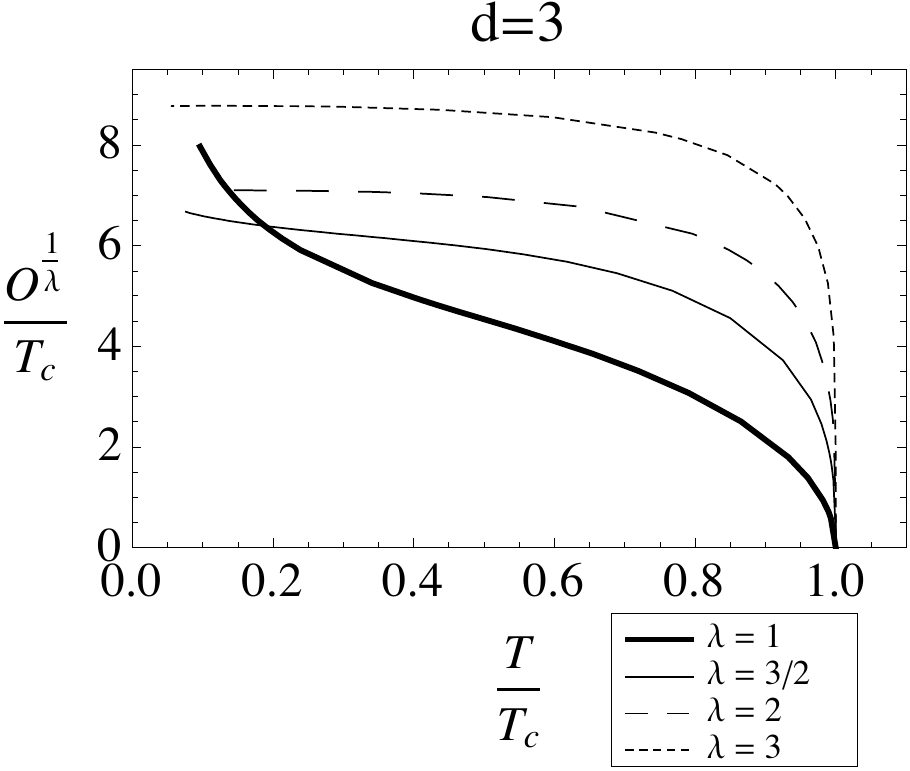}
\includegraphics[width=.45\textwidth]{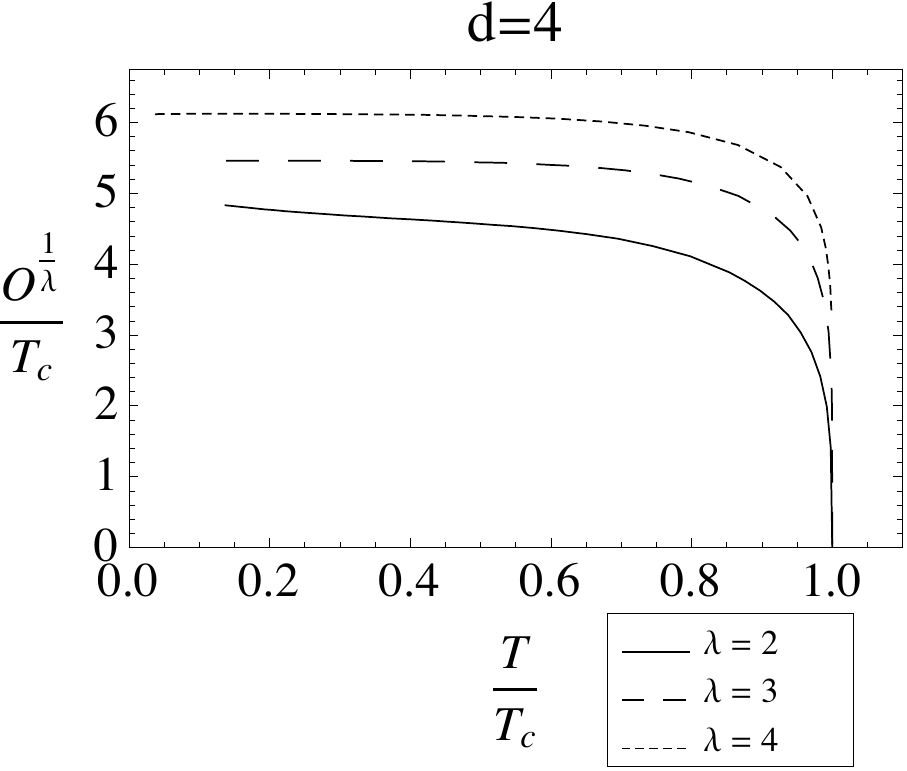}
\caption{The condensate as a function of temperature. $\lambda$ is the dimension of the operator $\Op$, and $d$ is the spacetime dimension of the superconductor. $\lambda_{BF} = -3/2$ for  $d=3$ and $\lambda_{BF} = -2 $ for $d=4$. The condensate tends to increase with $\lambda$.}\label{order}
\end{center}\end{figure}

In figure \ref{order} we plot the condensate $\langle\Op\rangle$  for various masses  and spacetime dimension. The previously discussed case $m^2 = -2$ is included for comparison. It is convenient to label the curves by the dimension of the operator $\lambda$ rather than the mass $m$ in order to distinguish the two possible quantizations when $m^2 = -2$ and $d=3$.   In all cases there is a critical temperature, $T_c$, above which there is no condensate. The critical temperature is proportional to $\rho^{1/2}$ in $d=3$ and $\rho^{1/3}$ in $d=4$, reflecting the different dimensions of the charge density in the two cases. The ratios $T_c/\rho^{1/2}$ and $T_c/\rho^{1/3}$ decrease as $\lambda$ increases. This is expected since increasing $\lambda$ corresponds to increasing the mass of the bulk scalar,  and hence making it harder for the scalar hair to form. As expected from mean field theory, $\langle\Op\rangle\sim (T_c-T)^{1/2}$ near the critical point. At low temperature, for $\lambda>\lambda_{BF}$, the condensate quickly saturates a fixed value which increases with $\lambda$. For $\lambda=\lambda_{BF}$, the vev approaches a fixed value roughly linearly, and for $\lambda<\lambda_{BF}$ it appears to diverge. 

\section{Conductivity}

To observe that our boundary theory is superconducting, we need to calculate the conductivity $\sigma$. This is related to the retarded current-current two-point function for our global $U(1)$ symmetry, $\sigma(\omega)=\frac{1}{i\omega}G^R(\omega,k=0)$. To do so we must calculate an electromagnetic perturbation on top of the hairy black hole. The linearized equation of motion for $\delta A=A_x(r)e^{-i\omega t+iky}~dx$ is
\be
A_x'' +\left(\frac{f'}{f}+\frac{d-3}{r}\right)A_x' + \left(\frac{\omega^2}{f^2}-\frac{k^2}{r^2f}-\frac{2\psi^2}{f}\right)A_x=0 \label{currenteom}\ee
This mode is not coupled to other linearized perturbations  and can be studied separately. We require $A_x \propto f^{-i\omega/dr_0}$ near $r=r_0$ corresponding to ingoing wave boundary conditions at the horizon.
The AdS/CFT dictionary tells us how to calculate the retarded current Greens function $G^R$ from our gauge field perturbation \cite{Son:2002sd}. The result is
\be G^R=-\lim_{r\rightarrow\infty}f r^{d-3}A_xA'_x\ee where $A_x$ is purely infalling at the horizon, and normalized to $A_x(r=\infty)=1$.

Let us work this out for the two cases of interest to us, $d=3$ and $4$. As can be verified by the large $r$ expansion of (\ref{currenteom}), in $d=3$, the gauge field falls off as 
\be A_x=A^{(0)}+{A^{(1)}\over r}+\cdots\ee
 giving
\be G^R=\frac{A^{(1)}}{A^{(0)}}, \qquad \sigma(\omega)=\frac{1}{i\omega}G^R(\omega,k=0)=\frac{A^{(1)}}{i\omega A^{(0)}}|_{k=0}.\ee

The $d=4$ case is a little more nuanced. There is a logarithmic term in the falloff at nonzero frequency and momentum,
\be A_x=A^{(0)}+{A^{(2)}\over r^2}+\frac{A^{(0)}(\omega^2-k^2)}{2}\frac{\log \Lambda r}{r^2}\cdots \ee
giving
\be G^R=2\frac{A^{(2)}}{A^{(0)}}+(\omega^2-k^2)(\log\Lambda r-\frac{1}{2}).\ee
This logarithmic divergence can be removed with a boundary counterterm in the gravity action \cite{Taylor:2000xw}, but it necessarily breaks conformal invariance. In other words, we must specify a renormalization scale when regulating the action. After adding a counterterm to cancel the $\log \Lambda r$ term we have the Greens function
\be G^R=2\frac{A^{(2)}}{A^{(0)}}+\frac{k^2-\omega^2}{2}, \qquad \sigma=\frac{2A^{(2)}}{i\omega A^{(0)}}|_{k=0} +\frac{i\omega}{2}.
\ee
It is important to note that when we are calculating physical quantities such as $n_n,~n_s,~\Delta,~\omega_g$, these will all be invariant under a shift in $\Lambda$, as all this can change is the quadratic term in $\Re[G^R]$. It will however effect the definition of our correlation length, as will be explained later.

\begin{figure}
\begin{center}
\includegraphics[width=.45\textwidth]{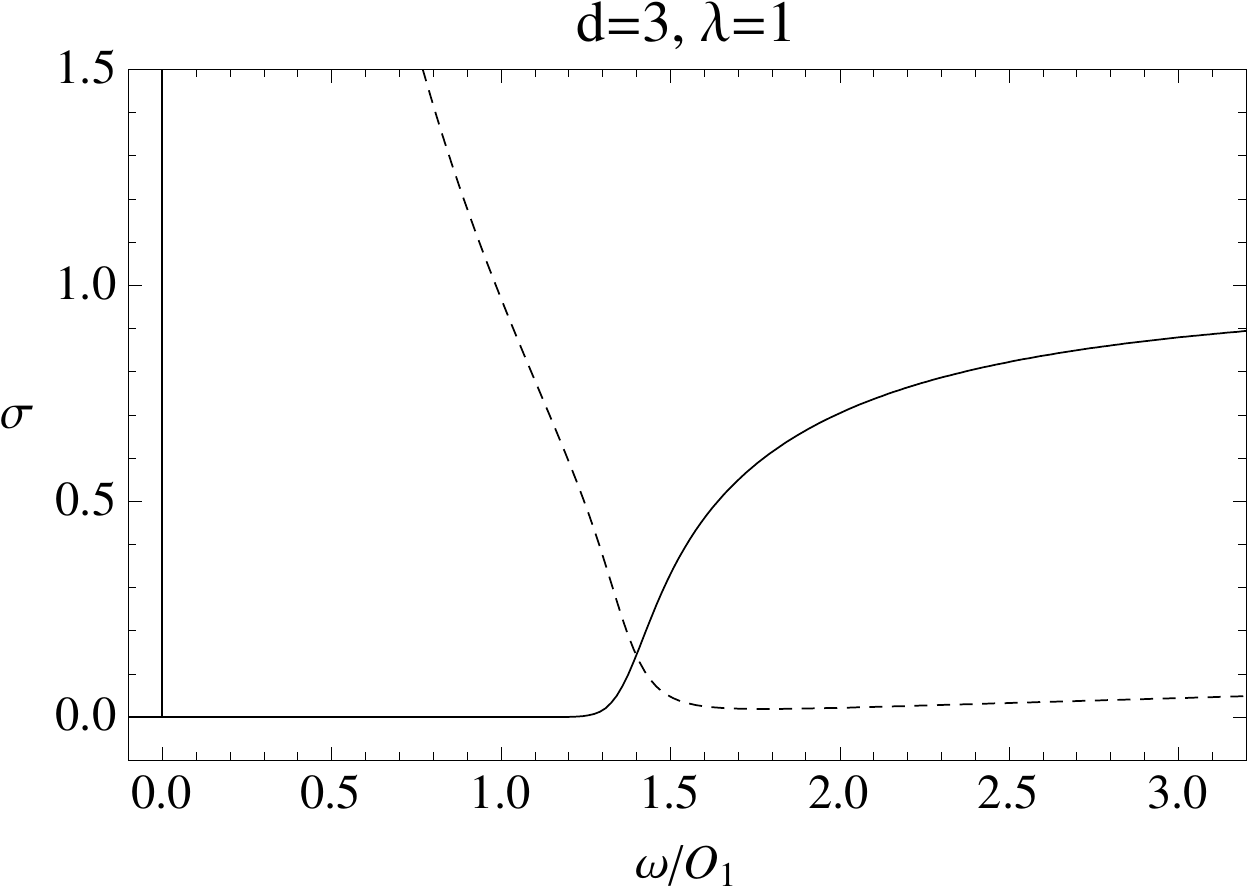}\hspace{0.2cm}
\includegraphics[width=.45\textwidth]{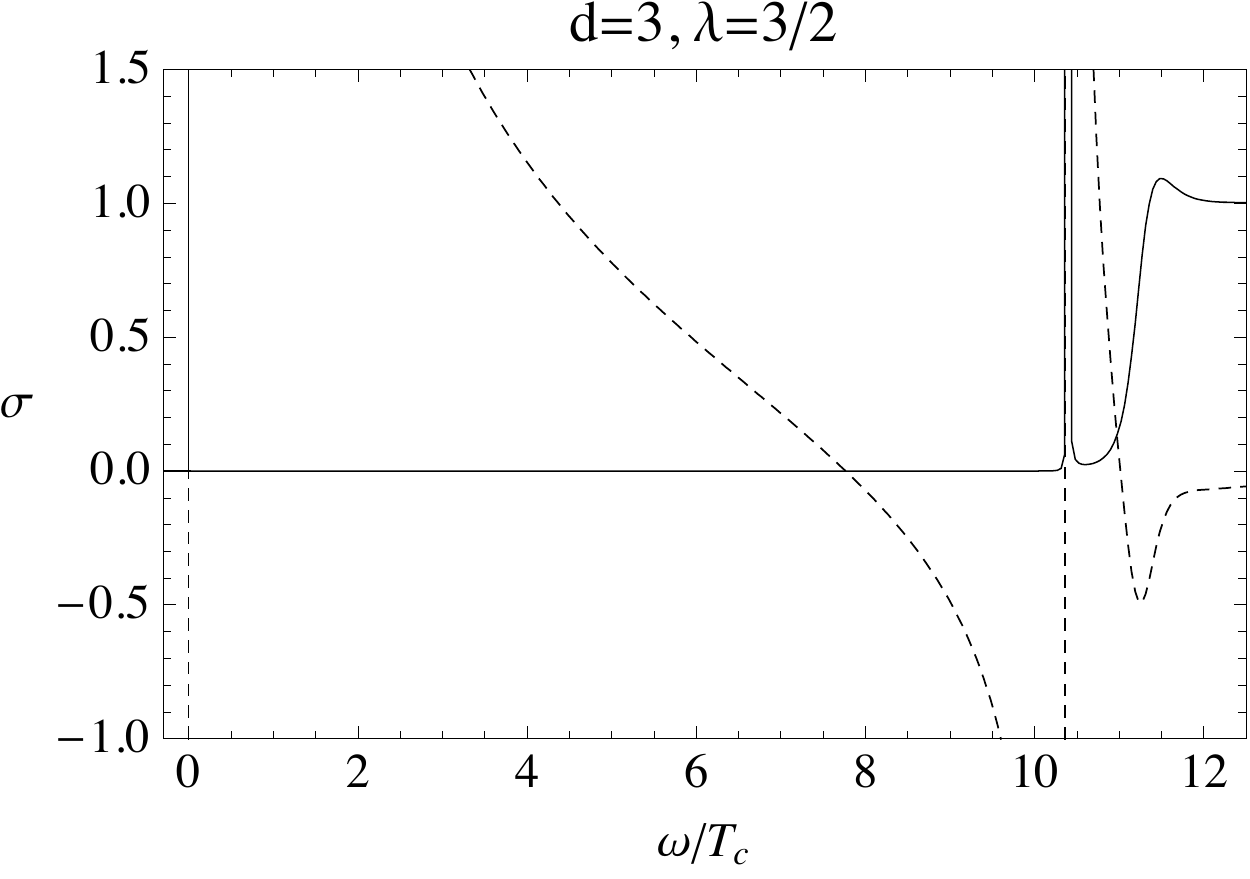}\\
\vspace{1cm}
\includegraphics[width=.45\textwidth]{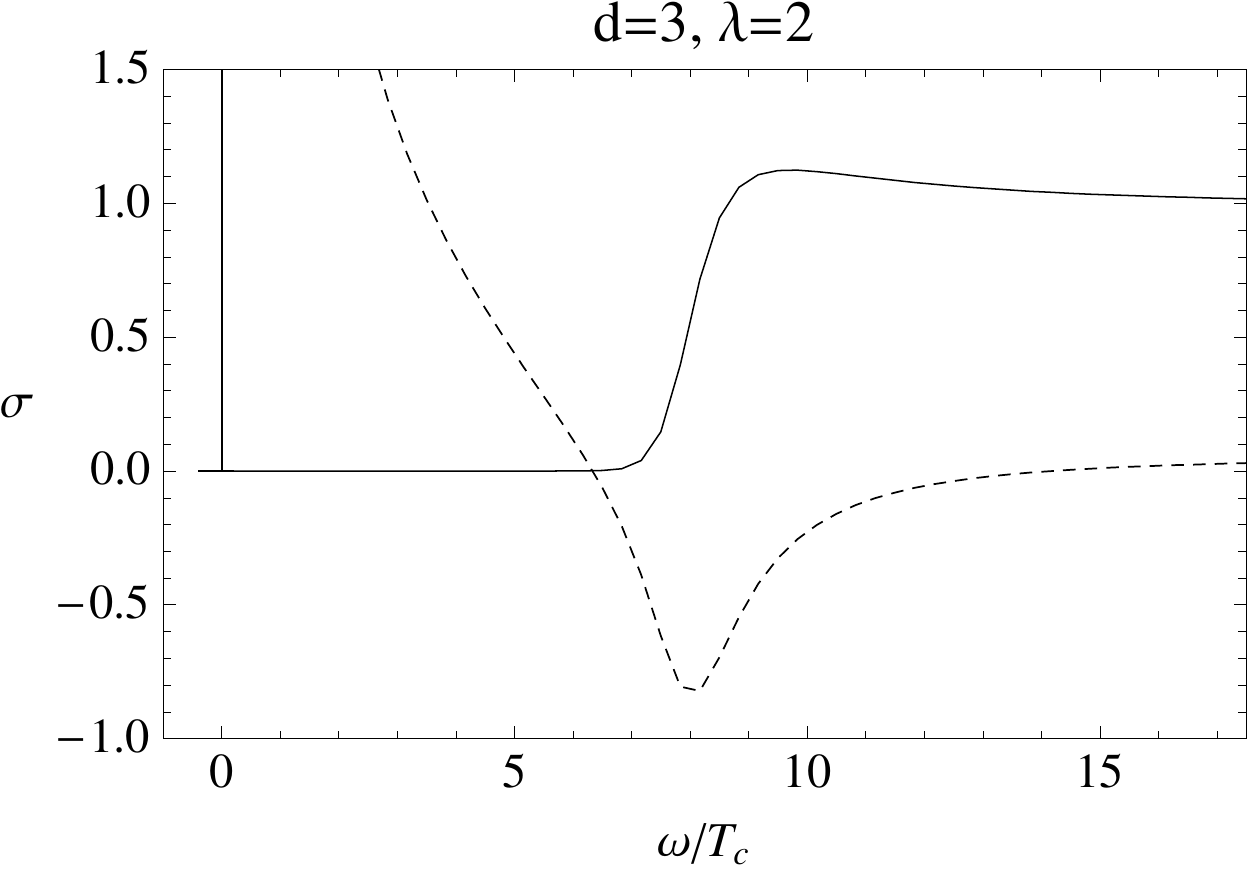}\hspace{0.2cm}
\includegraphics[width=.45\textwidth]{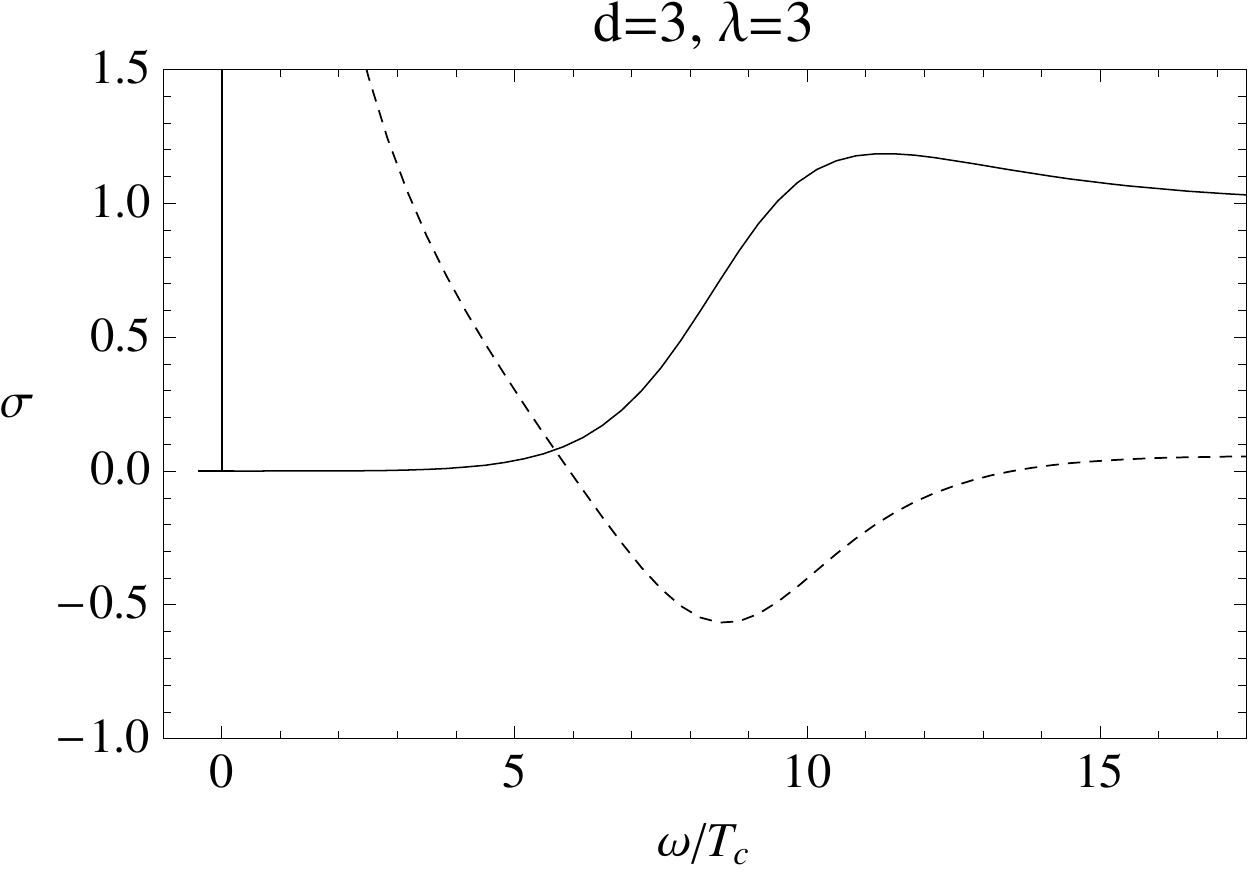}
\caption{Conductivity for $2+1$ dimensional superconductors. Each plot is at low temperatures, about $T/T_c\approx 0.1$. The solid line is the real part, dashed is imaginary. The pole at $\omega =0 $ is clearly visible in $\Im[\sigma]$. For $\lambda\ge 3/2$, we find the shape of the curve near the edge of the gap largely dictated by a second pole in the lower half complex $\omega$ plane. As $T\rightarrow0$ and $\lambda\rightarrow 3/2$, this pole hits the real axis.}\label{d3cond}
\end{center}\end{figure}

\begin{figure}\begin{center}
\includegraphics[width=.45\textwidth]{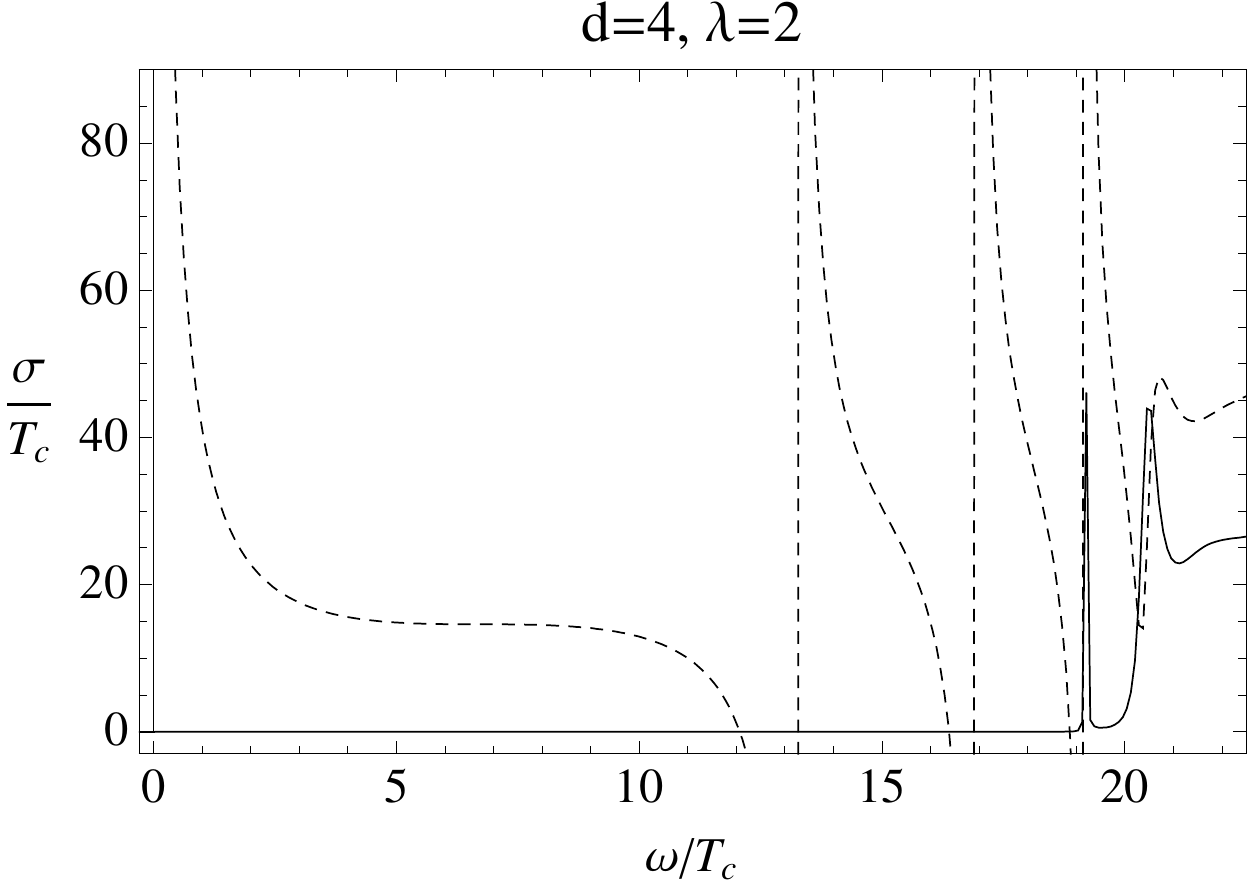}\\
\vspace{1cm}
\includegraphics[width=.45\textwidth]{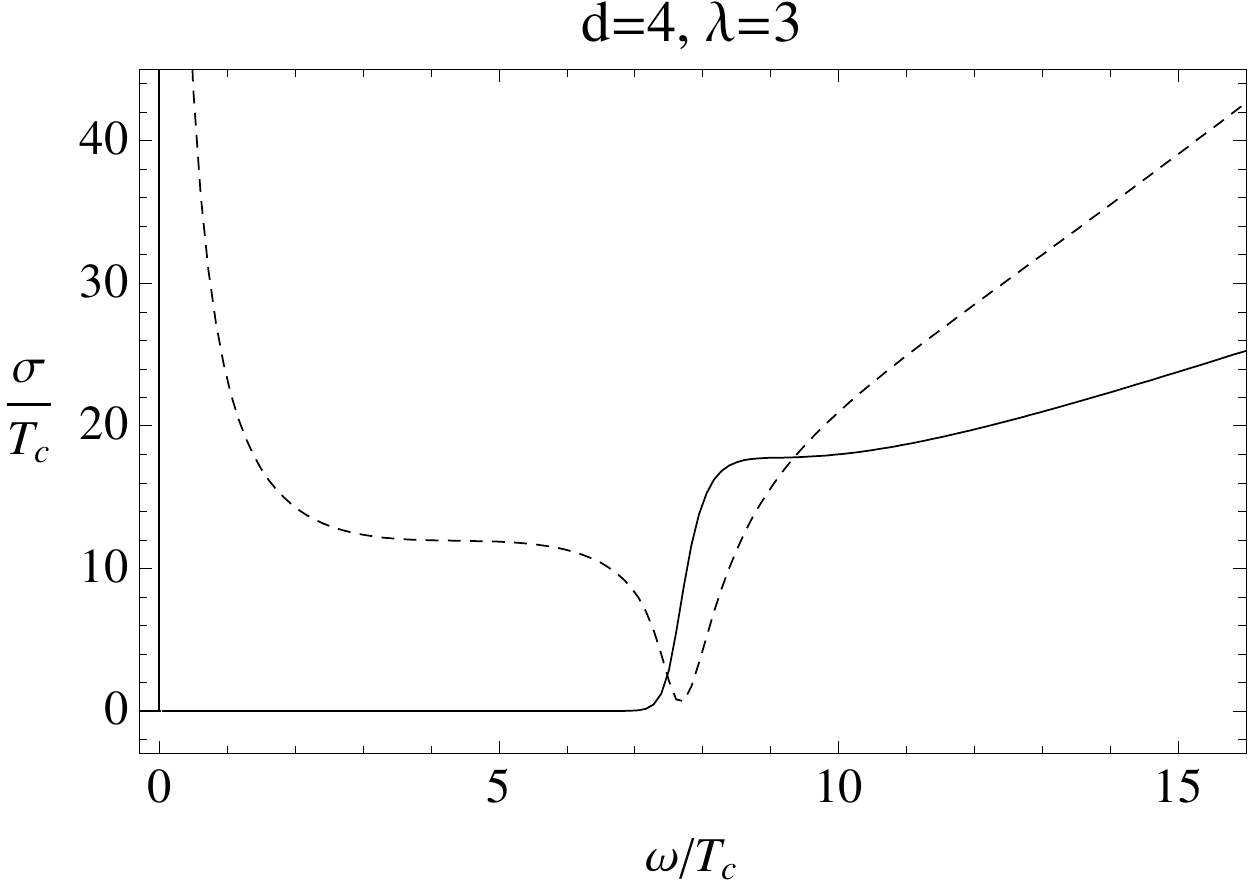}\hspace{0.2cm}
\includegraphics[width=.45\textwidth]{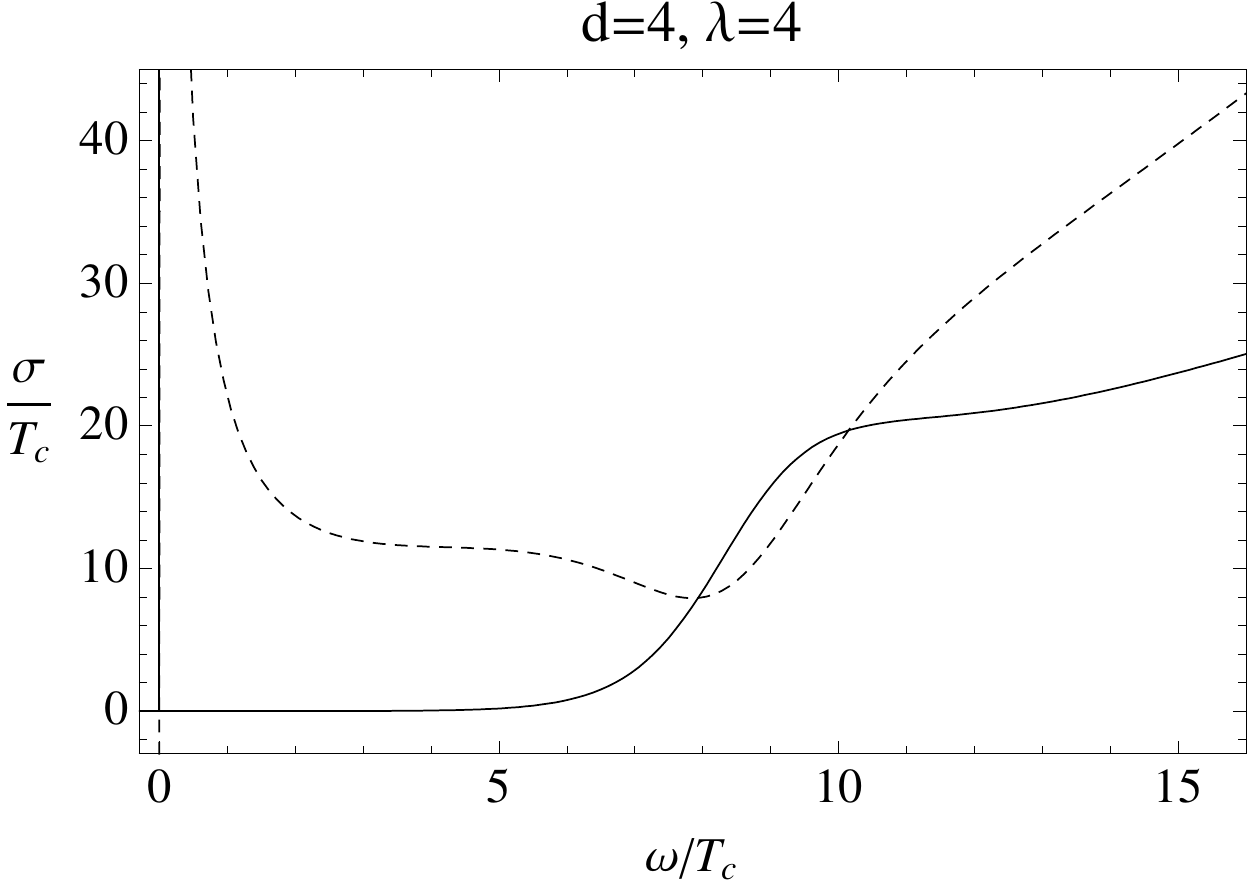}
\caption{Conductivity for $3+1$ dimensional superconductors. Each plot is at low temperatures, about $T/T_c\approx 0.1$. The solid line is the real part, dashed is imaginary. The pole at $\omega =0 $ is clearly visible in $\Im[\sigma]$. For $\lambda= 2$, we find the shape of the curve largely dictated by multiple poles. As $T\rightarrow0$ and $\lambda\rightarrow 2$, these poles approach and hit the real axis.}\label{d4cond}
\end{center}\end{figure}

In Figures \ref{d3cond} and \ref{d4cond} we plot the conductivity $\sigma(\omega)$ for the various cases studied. It is clear that in all cases we find a gap in the conductivity. We parameterize this by the gap frequency $\omega_g$. At zero temperature we expect there to be a sharp gap, indicating that there are no states below said energy scale. In most cases the conductivity rises quickly near $\omega_g$, the exception being the case when the bulk scalar is massless. In this case, while there is exponential suppression in $\sigma$ at $\omega\ll\omega_g$, the conductivity rises more gradually around the gap frequency even at low temperature.

 At nonzero temperature, 
there is some ambiguity in defining $\omega_g$ when looking at the thermally smoothed conductivity plots. However, it turns out that  for $\lambda > \lambda_{BF}$, $\Im[\sigma]$ always has a minimum right in the region where $\Re[\sigma]$ is turning on. We will define $\omega_g$ as the frequency minimizing $\Im[\sigma]$. For $\lambda < \lambda_{BF}$,   $\Im[\sigma]$ decreases monotonically and we define $\omega_g$ by the minimum of $|\sigma|$. 

We find that $\omega_g$ remains quite uniform for many of the cases we have studied. For all cases with $\lambda>\lambda_{BF}$, we find 
\be\label{omtc}
{\omega_g\over T_c}\approx 8
\ee
with deviations of less than 8\%. Even changing the spacetime dimension of the superconductor from $d=3$ to $d=4$ does not affect this ratio. Since the corresponding BCS value is 3.5, this shows that the energy to break apart the condensate is more than twice the weakly coupled value. 

The most interesting case is $\lambda =\lambda_{BF}$. Here we find that additional poles in $\Im[\sigma]$ and delta functions in $\Re[\sigma]$ arise at low temperature inside the gap.   One can interpret such poles as follows. In standard BCS theory, the quasiparticles excited above the superconducting ground state are not interacting. Since holographic superconductors are strongly coupled, it is perhaps not surprising to find interactions between the quasiparticles.  For $\lambda > \lambda_{BF}$, the Greens function $G^R$ has poles in the complex frequency plane which produce  the bump in $\Re[\sigma]$ just above the gap. This looks like  a resonance in quasiparticle scattering. As $\lambda \rightarrow \lambda_{BF}$ and $T \rightarrow 0$,  the pole moves onto the real axis, indicating that the interactions become stronger and form a bound state. For $d=3$, there is one such pole, and the difference $\omega_g - \omega_{pole}$ reflects the binding energy of the bound state. For $d=4$, more poles appear as one lowers the temperature. Since our numerics can only explore down to $T/T_c \approx 0.1$ it is not clear what the total number is, and what the ultimate gap $\omega_g$ will be. It would be  very interesting to study these cases with backreaction to see how much of this pole structure remains.

On the gravity side these poles are also surprising, as they imply that our near extremal black hole has normal modes. More precisely, at the frequency $\omega_{pole}$ there are solutions for a vector potential $A_x$ which are ingoing at the horizon and vanish at infinity. Usually, black holes have quasinormal modes with complex frequencies that describe the decay of perturbations outside the horizon. In AdS, these correspond to poles in the retarded Greens function \cite{Son:2002sd,Kovtun:2005ev}.  Remarkably, when $m^2 = m^2_{BF}$ and $T \rightarrow 0$, some quasinormal frequencies become real. There are vector perturbations which do not decay. We do not have an intuitive explanation for this. It would be interesting to understand why this is happening and check if there are similar normal modes associated with the scalar field $\psi$ as well. 

\begin{figure}\begin{center}
\includegraphics[width=.45\textwidth]{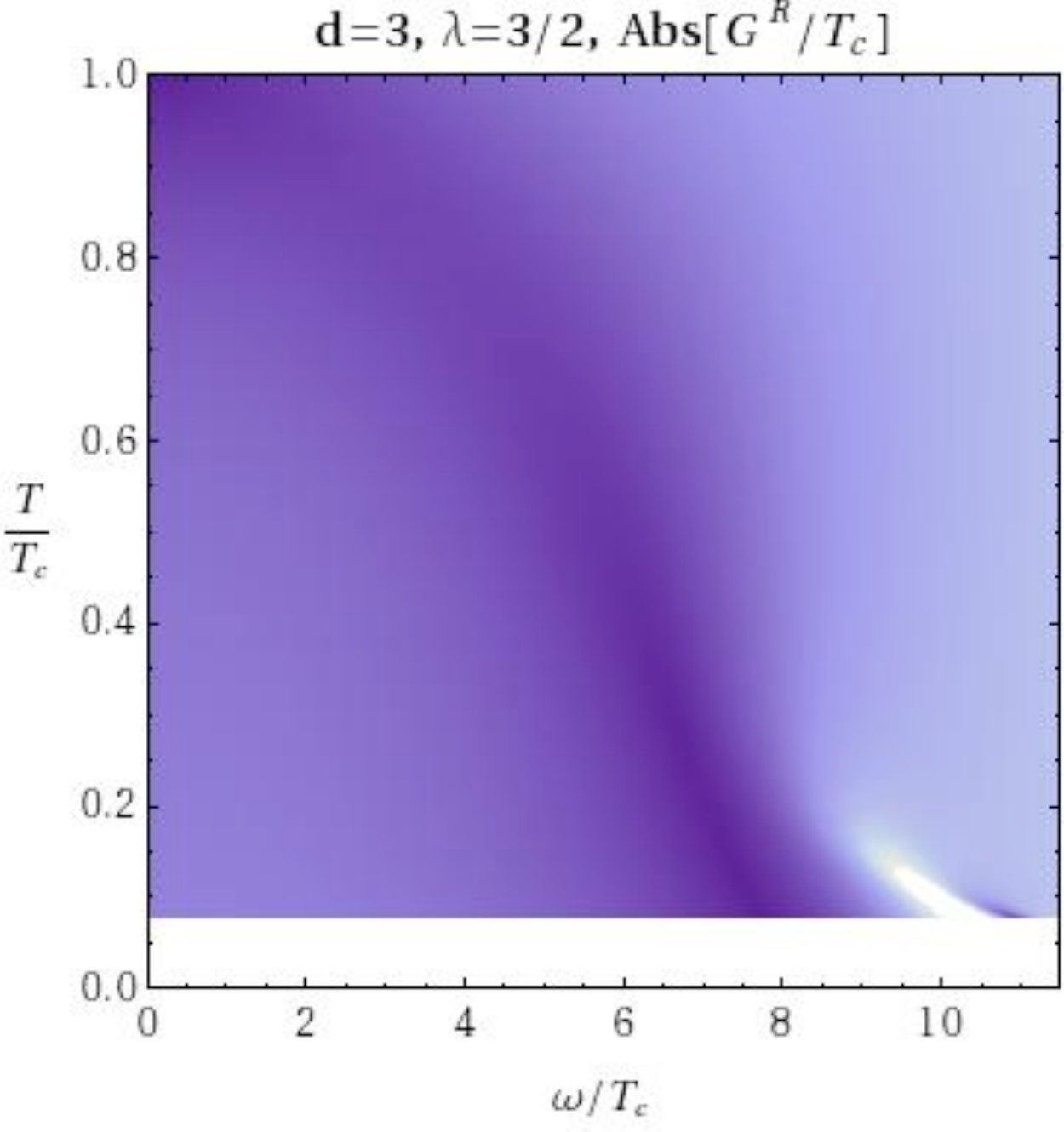}
\includegraphics[width=.45\textwidth]{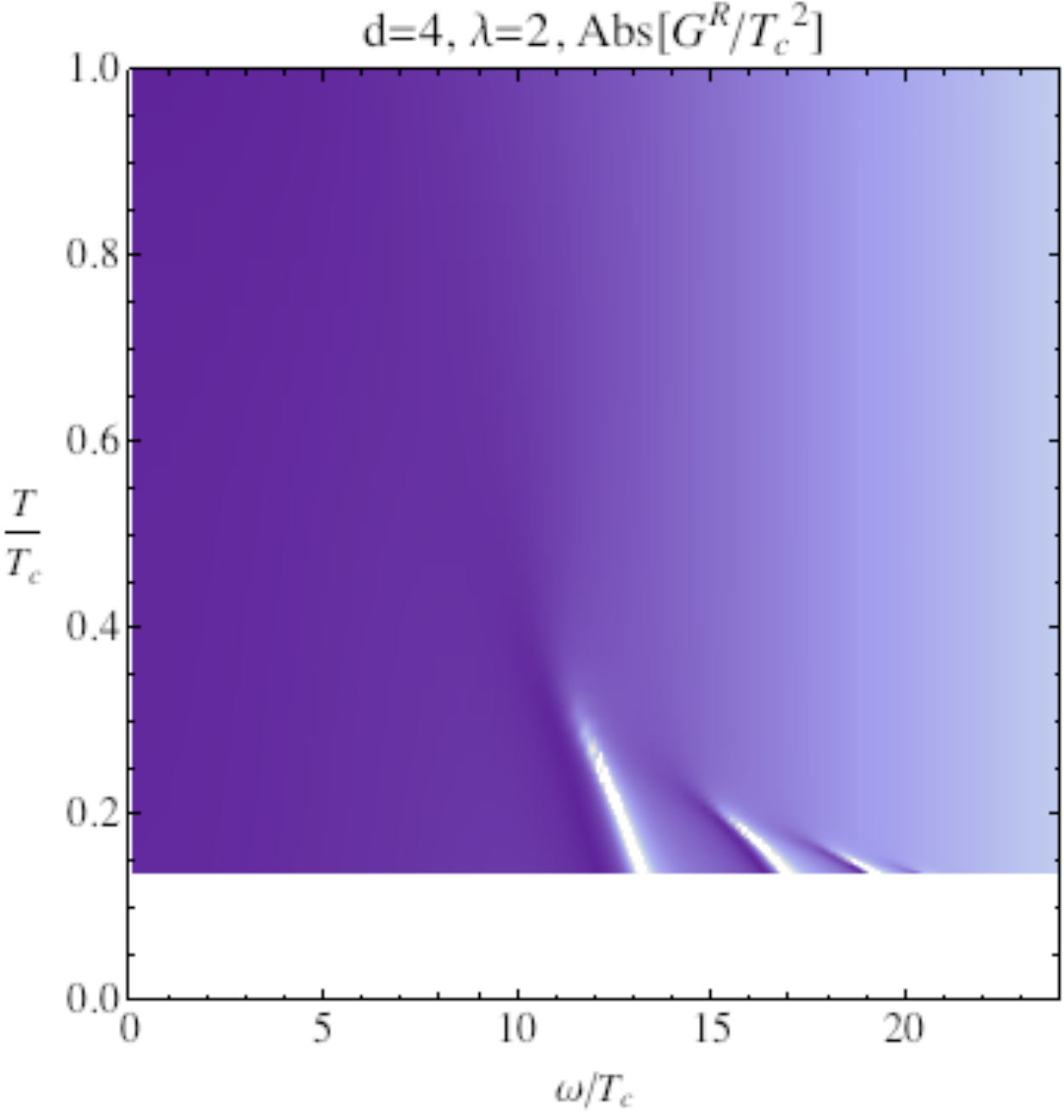}
\caption{The absolute value of the retarded Greens function is shown as a function of temperature and (real) frequency for $\lambda=\lambda_{BF}$ at zero spacial momentum. Lighter shades denote larger values. One can clearly see the complex frequency poles as they move onto the real axis.}\label{conddensity}
\end{center}\end{figure}

Figure \ref{conddensity} shows $|G^R|$ as a function of temperature and (real) frequency for the $\lambda =\lambda_{BF}$ cases. Lighter shades denote larger values. One can clearly see the poles as they move onto the real $\omega$ axis. One can infer the conductivity at different temperatures by looking at horizontal lines in this diagram.

\section{Other properties of the superconductor}

In BCS theory, one often studies the order parameter  $\Delta(T)$, the mass of charged quasiparticle excitations about the fermi surface as a function of temperature. However, the gravity dual does not have direct access to the elementary charged particles in the dual superconductor, so we cannot calculate this in general. We can, however, extract the mass of the low lying excitations at low temperature. We look at the normal contribution to the DC conductivity, $n_n=\lim_{\omega\rightarrow 0}\Re[\sigma]$, which will be exponentially suppressed when these modes are gapped, $n_n\sim e^{-\Delta(0)/T}$. This gives us the capability to numerically calculate $\Delta(0)$. Note that for $d=4$ the conductivity is dimensionful, and we read off $\Delta$ from $\sigma/T_c$.  In the BCS theory, $\omega_g=2\Delta$, due to the fact that the electron-photon coupling implies that a photon will excite quasiparticles in pairs. We find that this BCS result does not hold in general, even though it did hold for the cases studied in  \cite{Hartnoll:2008vx}. The values of $\Delta$ are given in Tables 1 and 2 at the end of this paper.

Another order parameter is the superfluid density, $n_s$. It is the coefficient of the pole in  $\Im[\sigma]$ at $\omega=0$. By the Kramers-Kronig relations, if $\Im[\sigma]=n_s/\omega$, $\Re[\sigma]=\pi n_s \delta(\omega)$. It is this delta function that tells us our boundary theory exhibits DC superconductivity. The BCS theory gives  relations between $n_s,~\langle\Op\rangle$, and $\Delta$, but for holographic superconductors,  no analogous formulas are known. 

\begin{figure}\begin{center}
\includegraphics[width=.45\textwidth]{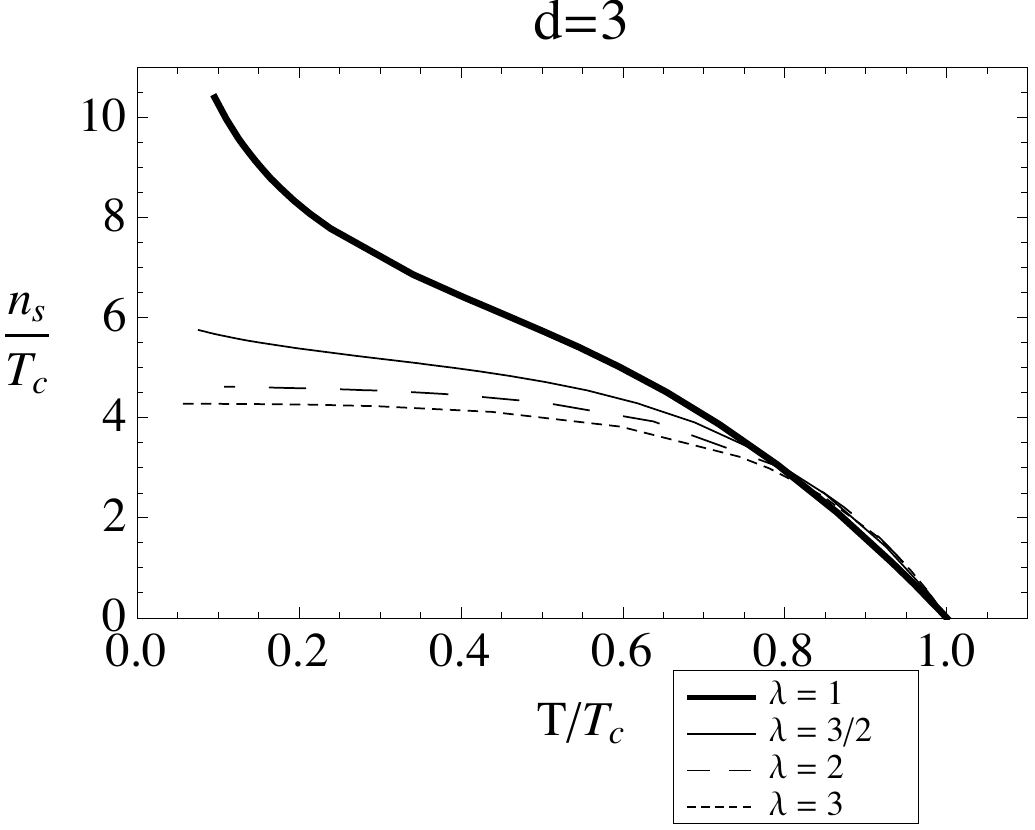}
\includegraphics[width=.45\textwidth]{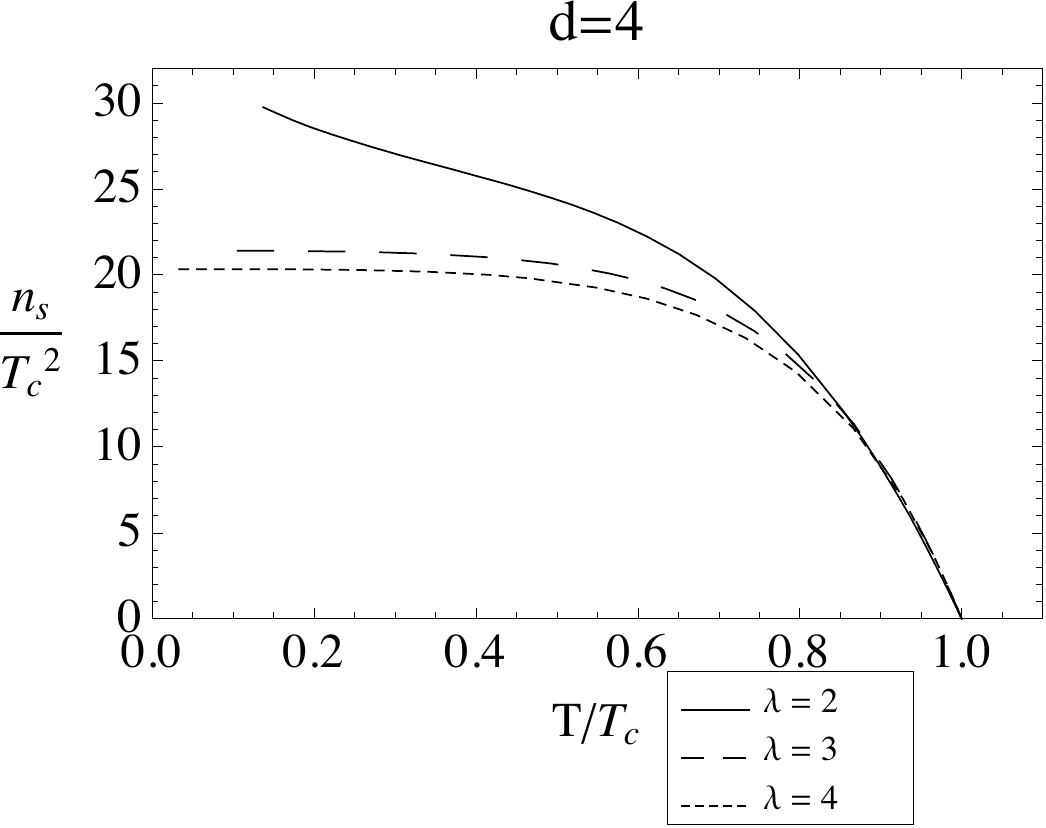}
\caption{Superfluid density in $d=3$ and $d=4$.}\label{nsplots}
\end{center}\end{figure}

It is clear from the definition of $\sigma$ that $n_s=\Re[G^R(\omega=k=0)]$. In terms of long-wavelength response, we can extract two correlation lengths, $\xi_\omega$ and $\xi_k$. We simply study the small $\omega,~k$ behavior of $G^R$,
\be\Re[G^R]=n_s(1-\xi_\omega^2\omega^2+\xi_k^2 k^2+\cdots)\ee
Because Lorentz invariance is broken at finite temperature, we do not expect $\xi_\omega=\xi_k$. However, at low temperatures the spacetime is in some sense ``nearly''  $AdS$ in Poincare coordinates, and so at low temperatures we expect the two values to be approximately equal. In the $d=3$ case we indeed find that they are nearly equal not only at low temperature, but all the way up to $T/T_c\approx0.7$. For $d=4$,  $\xi_\omega$ and $\xi_k$ are not separately well defined since there is a renormalization ambiguity as mentioned earlier. We can always shift $G^R\rightarrow G^R+C(\omega^2-k^2)$. Therefore in that case the only renormalization-independent quantity is the difference, $\delta\xi^2=\xi_k^2-\xi_\omega^2$. Studying $G^R$ at small $k$ can be understood as linear response to a long wavelength magnetic field, it is unclear how to interperet $\xi_\omega$.

The results for the superfluid density $n_s$ and correlation length are given in Figures 5 and 6. We find a surprising relation between  $n_s$ and the gap frequency $\omega_g$. 
For all $\lambda>\lambda_{BF}$ we find
\be\label{nsog}
 {n_s\over \omega_g}\approx \frac{1}{2} \quad    (d=3), \qquad 
   {n_s\over \omega_g^2} \approx \frac{1}{3} \quad  (d=4)
   \ee
The agreement is better than 10\%. 
For $d=3,~\lambda=1$ we find that $n_s/\omega_g = 1$.  This agrees with the analytic zero temperature approximate solution in \cite{us}. With $\psi(r)\approx\langle\Op_1\rangle/r$, one finds $G^R(T=0)=\sqrt{2\langle\Op_1\rangle^2-\omega^2+k^2}$. Note that this also gives the prediction $\xi_k=\frac{1}{2\langle\Op_1\rangle}=\frac{1}{\sqrt{2}\omega_g}$. We indeed find from numerics that $\xi_k \langle\Op_1\rangle=0.51$. 
 
We also find that, for all four values of $\lambda$ studied in $d=3$, 
    \be
    \xi_k(0)\langle\Op_\lambda\rangle^{1/\lambda}\approx .18+.37 \lambda\ .
    \ee
   It would be very interesting to see how much of this universal behavior remains when backreaction is included. As mentioned above,  we cannot calculate $\xi_k \langle\Op_\lambda\rangle^{1/\lambda}$ in $d=4$, and due to restoration of Lorentz invariance $\delta\xi$ vanishes at low temperatures. 
   
We  find universal behavior near the critical point. As was pointed out earlier,  the condensate vanishes as $(T_c-T)^{1/2}$, as expected from mean field theory. The superfluid density always vanishes linearly, as expected from Landau-Ginzburg theory, and the coefficient is roughly  the same  (within 15\%) for each dimension: 
\be
n_s\approx 20 (T_c-T) \quad  (d=3),  \qquad   n_s\approx 100 T_c(T_c-T) \quad  (d=4) \ .
\ee
Similarly, the correlation length diverges as $(T_c-T)^{-1/2}$ as is expected from Landau-Ginzburg with nearly exactly the same coefficient in each dimension:
\be\label{corlength}
\xi_k= \frac{0.1}{T_c}(1-T/T_c)^{-1/2} \quad (d=3), \qquad \xi_k= \frac{0.06}{T_c}(1-T/T_c)^{-1/2} \quad (d=4)\ .
\ee
 A  correlation length defined from perturbations in the $\psi$ field was found in the $\lambda=2$ case to have the same divergence in \cite{Maeda:2008ir}, and one would expect it to generalize to arbitrary $\lambda$. The difference is that the correlation length we calculate is similar to that defined in BCS theory, i.e.,  the small spatial momentum correction to the current-current correlator, whereas Maeda and Okamura calculate the Landau-Ginzburg correlation length, obtained from fluctuations of the charged scalar field.

\begin{figure}\begin{center}
\includegraphics[width=.45\textwidth]{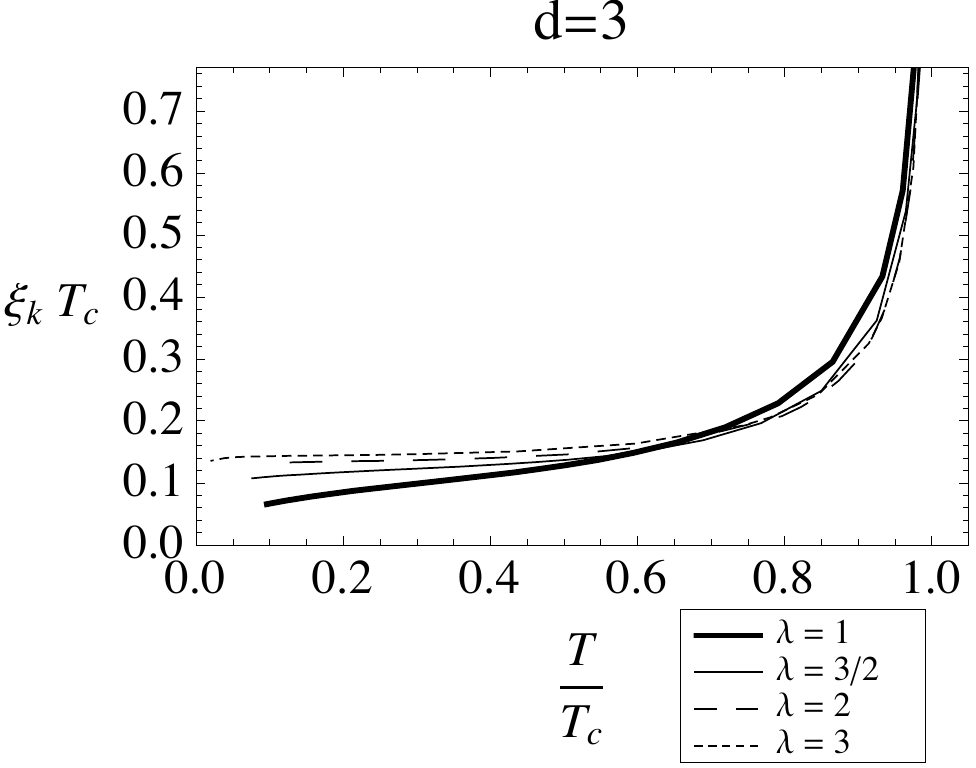}
\includegraphics[width=.45\textwidth]{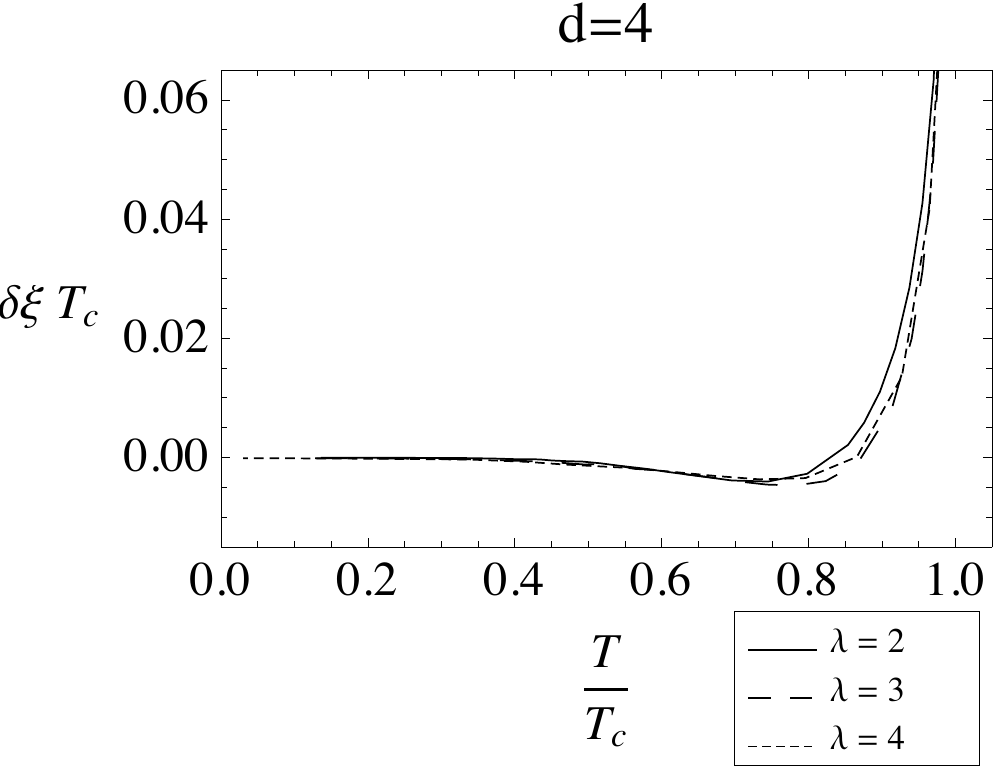}
\caption{Correlation length in $d=3$ and $d=4$.}\label{corrplots}
\end{center}\end{figure}

\section{Summary}

We have studied holographic superconductors in both $2+1$ and $3+1$ dimensions, with operators of different conformal weight condensing. One advantage of going to $3+1$ dimensions is that there are various subtleties in $2+1$ dimensions associated with the Mermin-Wagner theorem stating that one cannot break continuous symmetries and the existence of  Kosterlitz-Thouless transitions. The fact that we see qualitatively the same behavior in both cases shows that these subtleties do not play any role here. This is probably because we are using classical gravity to describe the superconductor, which is valid in a large $N$ limit.

A summary of our  numerical results are enumerated in Table 1 and Table 2 below. There is no entry for $\omega_g$ in the $\lambda=\lambda_{BF}$ cases because, as we have seen, additional poles arise as $T\rightarrow 0$ and $\omega_g$ increases. At the smallest $T/T_c$ that we can explore, the conductivity has not reached a limiting curve. The value of $\Delta$ seems anomalously low for $d=3$, $\lambda = 3$. This is a case where the conductivity rises slowly near $\omega_g$ even at low $T$ and it is possible that this has contaminated our measurement of $\Delta$.

We have found a number of surprising regularities involving the gap frequency $\omega_g$, the critical temperature $T_c$, the superfluid density $n_s$ and the correlation length $\xi_k$, see eq. (\ref{omtc}) and (\ref{nsog}) - (\ref{corlength}). Some of these are clearly visible in the tables, such as the last line, and the row for $\omega_g$. Others involve comparing two quantities. It would be interesting to investigate this further and determine  if some of these regularities lead to universal predictions for holographic superconductors.

\begin{table}[htdp]
\begin{center}\begin{tabular}{|c|cccc|}\hline
$ \lambda$ &1& 3/2 & 2 & 3\\ \hline
 $m^2$ &-2&  -9/4 & -2 & 0 \\
 $T_c$ &$0.226 \sqrt{\rho}$& $0.152 \sqrt{\rho}$ & $0.118 \sqrt{\rho}$ &$0.087 \sqrt{\rho}$ \\
 $\langle\Op_\lambda\rangle^{1/\lambda}(0) $&divergent  &$7.0 T_c$ &$7.1T_c$ &$8.8T_c$ \\
 $\langle\Op_\lambda\rangle^{1/\lambda}(T\approx T_c) $&$7.0T_c(1-t)^{1/2\lambda} $ &$9.1T_c(1-t)^{1/2\lambda}$ &$10.0T_c(1-t)^{1/2\lambda}$  &$11.0T_c(1-t)^{1/2\lambda}$ \\
$ \omega_g$&$1.4\langle\Op_1\rangle$ & & $8.1T_c$ & $8.6T_c$ \\
 $\Delta$ &$0.8\langle\Op_1\rangle$ & $5.3T_c$  & $4.2T_c$  & $0.6T_c$ \\
$ n_s(0) $& $1.3\langle\Op_1\rangle$&$6.0T_c$& $4.7T_c$ & $4.3T_c$ \\
$ n_s(T\approx T_c) $&$17.0(T_c-T)$  &$20.9(T_c-T)$ & $23.5(T_c-T)$ & $21.4(T_c-T)$ \\
$\xi_k(0) $& $0.55/\langle\Op_1\rangle$&$0.10/T_c$ &$ 0.13/T_c$ & $0.14/T_c$ \\
$\xi_k(T\approx T_c) $&$\frac{0.11}{T_c}(1-t)^{-1/2}$ &$\frac{0.11}{T_c}(1-t)^{-1/2}$ &$ \frac{0.10}{T_c}(1-t)^{-1/2}$&  $\frac{0.10}{T_c}(1-t)^{-1/2}$ \\ 
 \hline \end{tabular} \caption{Various $2+1$ superconductors. There is no entry for $\omega_g$ when $\lambda = 3/2$ due to complications arising from complex poles moving onto the real axis (see text). Note $t=T/T_c$.}
\end{center}
\label{3dtable}
\end{table}

\begin{table}[htdp]
\begin{center}\begin{tabular}{|c|ccc|}\hline
$ \lambda$ &2& 3& 4 \\ \hline
 $m^2$ &-4&  -3 & 0 \\
 $T_c$ &$0.253 \rho^{1/3}$& $0.198\rho^{1/3}$ & $0.170\rho^{1/3}$ \\
 $\langle\Op\rangle^{1/\lambda}(0) $& $5.0T_c$ &$5.5T_c $& $6.1T_c$ \\
 $\langle\Op\rangle^{1/\lambda}(T\approx T_c) $&$6.7T_c(1-t)^{1/2\lambda} $ &$7.3T_c(1-t)^{1/2\lambda}$ &$7.7T_c(1-t)^{1/2\lambda}$ \\
$ \omega_g$& & $7.7T_c$& $7.8T_c$ \\
 $\Delta$ &$8.6T_c$ & $5.5T_c$  & $4.2T_c$\\
$ n_s(0) $& $32.5T_c^2$ &$21.4T_c^2$  & $20.3T_c^2$\\
$ n_s(T\approx T_c) $&$97.4T_c^2(1-t)$  &$104T_c^2(1-t)$  &$111T_c^2(1-t)$  \\
$ \delta\xi(T\approx T_c) $& $\frac{0.06}{T_c}(1-t)^{-1/2}$ &$\frac{0.06}{T_c}(1-t)^{-1/2}$ &$\frac{0.06}{T_c}(1-t)^{-1/2}$  \\
 \hline \end{tabular} \caption{Various $3+1$ superconductors. There is no entry for $\omega_g$ when $\lambda = 2$ due to complications arising from complex poles moving onto the real axis (see text). Note $t=T/T_c$.}
\end{center}
\label{defaulttable}
\end{table}

\vskip 1cm
\centerline{\bf Acknowledgements}
\vskip .5cm
It is a pleasure to thank Matthew Fisher, Andreas Ludwig, Sean Hartnoll and Don Marolf for stimulating discussions. This work was supported in part by NSF grant PHY-0555669.

\appendix
\section{Normal phase solution}

In this appendix, we calculate the conductivity in the normal phase, when the condensate vanishes, for both $d=3$ and $d=4$. It turns out that this can be done analytically.
At zero spatial momentum with no scalar hair, one can solve the wave equation for $A_x$ exactly. The infalling solutions are
\be\label{3dexact} A_x(r)= \exp\left[- \frac{i\omega}{6 r_0}\left(\log\frac{(r-r_0)^2}{r^2+r r_0+r_0^2} +2\sqrt{3}\arctan\frac{2r+r_0}{\sqrt{3}r_0}  \right) \right] (d=3)
\ee
\be A_x(r)=\frac{((\frac{r}{r_0})^2-1)^{- i\omega/4r_0}}{(1+(\frac{r}{r_0})^2)^{\omega/4r_0}}{}_2F_1[-(\frac{1}{4}+\frac{i}{4})\frac{\omega}{r_0},1-(\frac{1}{4}+\frac{i}{4})\frac{\omega}{r_0},1-\frac{i\omega}{2r_0},(1-(\frac{r}{r_0})^2)/2] (d=4)
\label{4dexact}\ee
Expanding at large $r$ yields the normal phase
 conductivities
\be \sigma^{d=3}_n=1\ee
\be \sigma^{d=4}_n=i\omega \left[\frac{1}{2}\psi\left(\frac{(1-i)\omega}{4\pi T}\right)+\frac{1}{2}\psi\left(-\frac{(1+i)\omega}{4\pi T}\right) +\log \frac{2^{1/2}\pi T}{T_c}+\gamma\right]-\pi T,\label{exactcond}\ee
where here $\psi(z)$ is the digamma function $\psi(z)=\Gamma'(z)/\Gamma(z)$.

For $\omega\ll T$,
\be \sigma^{d=4}_n=T\left(\pi  +i\log\left(\frac{\sqrt{2}\pi T}{T_c}\right)\frac{\omega}{T}+\frac{\pi\omega^2}{24 T^2}\cdots\right)\ee
and for $\omega\gg T$,
\be  \sigma^{d=4}_n=\omega\left( \frac{\pi}{2}+i(\log\frac{\omega}{2 T_c}+\gamma)-\frac{8 i \pi^4 T^4}{15 \omega^4}+\cdots\right). \ee

It is clear from (\ref{3dexact}) that  this particular perturbation has no homogenous quasinormal modes in $AdS_4$, though they exist at finite spatial momentum, as observed in \cite{Cardoso:2001vs}. This is due to the fact that the wave equation for a massless p-form field in $d+1$ dimensional planar AdS-Schwarzschild reduces to a one-dimensional wave equation which has a flat potential when $d=1+2p$ and $k=0$. In particular, (\ref{3dexact}) can be written $A_x(r) = e^{-i\omega r_*}$ where $r_* = \int dr/f$.

The $AdS_5$ case has quasinormal modes at
 \be\omega_{QNM}=2\pi T(1-i)n,~n=1,2,\cdots\ee
As was noted in \cite{Kovtun:2005ev}.

Our numerical results agree with these analytic formula for $T>T_c$, as well as $T<T_c$ and $\omega\gg\omega_g$. This is expected, as at large $\omega$  the frequency term dominates over the scalar hair contribution in the wave equation (\ref{currenteom}).

\end{document}